\newcommand{\CK}{\v{C}erenkov}
\documentclass{ws-procs9x6} % Computer Modern font calls

\usepackage{overpic}
\usepackage{epsfig}
\begin{document}
\vspace{-2.cm}
\title
{Isotopic mass separation with the RICH detector of the AMS Experiment}
\vspace{-0.3cm}

\author{\underline{Lu\'isa Arruda}, F.Bar\~ao, J.Borges, F.Carmo,\\
  P.Gon\c{c}alves, R.Pereira M.Pimenta}
\address{LIP/IST \\
         Av. Elias Garcia, 14, 1$^o$ andar\\
         1000-149 Lisboa, Portugal \\
         e-mail: luisa@lip.pt}
\author{A. Keating}
\address{ESTEC/ESA, Netherlands}
\maketitle

%\vspace{-0.5cm}
\abstracts{ 
The Alpha Magnetic Spectrometer (AMS) to be installed on the 
International Space Station (ISS) will be equipped with a 
proximity focusing Ring Imaging \v Cerenkov detector (RICH).
Reconstruction of the \CK\ angle and the electric
charge with RICH are discussed. A likelihood method for the \CK\ angle
reconstruction was applied leading to a velocity determination for protons
with a resolution around 0.1\%. The electric charge reconstruction is based on the counting
of the number of photoelectrons and on an overall efficiency estimation on an
event-by-event basis. The isotopic mass separation of helium and beryllium
is presented.} 

%\vspace{-0.3cm}
\section{The AMS02 and the RICH detector}
AMS (Alpha Magnetic Spectrometer) [\refcite{bib:Ahlen}, \refcite{bib:Balebanov}] is a precision spectrometer 
designed to search for cosmic antimatter, dark matter and to study the relative abundances of elements and isotopic composition of the primary cosmic rays. 
It will be installed in the International Space Station (ISS), in 2008, where
it will operate for a period of at least three years.
It will be equipped with a Ring Imaging \CK\ detector (RICH). This detector was designed
to measure the velocity of singly charged particles with a resolution $\Delta
\beta$/$\beta$ of 0.1\%, to extend the electric charge separation up to the
iron element, to contribute to the albedo rejection and to contribute to the e/p separation.

\begin{figure}[htb]
\begin{center}
%\vspace{-.5cm}
\begin{tabular}{cc}  
\hspace{-.68cm}
\includegraphics[scale=0.3,angle=0,clip=,bb=14 14 599 396]{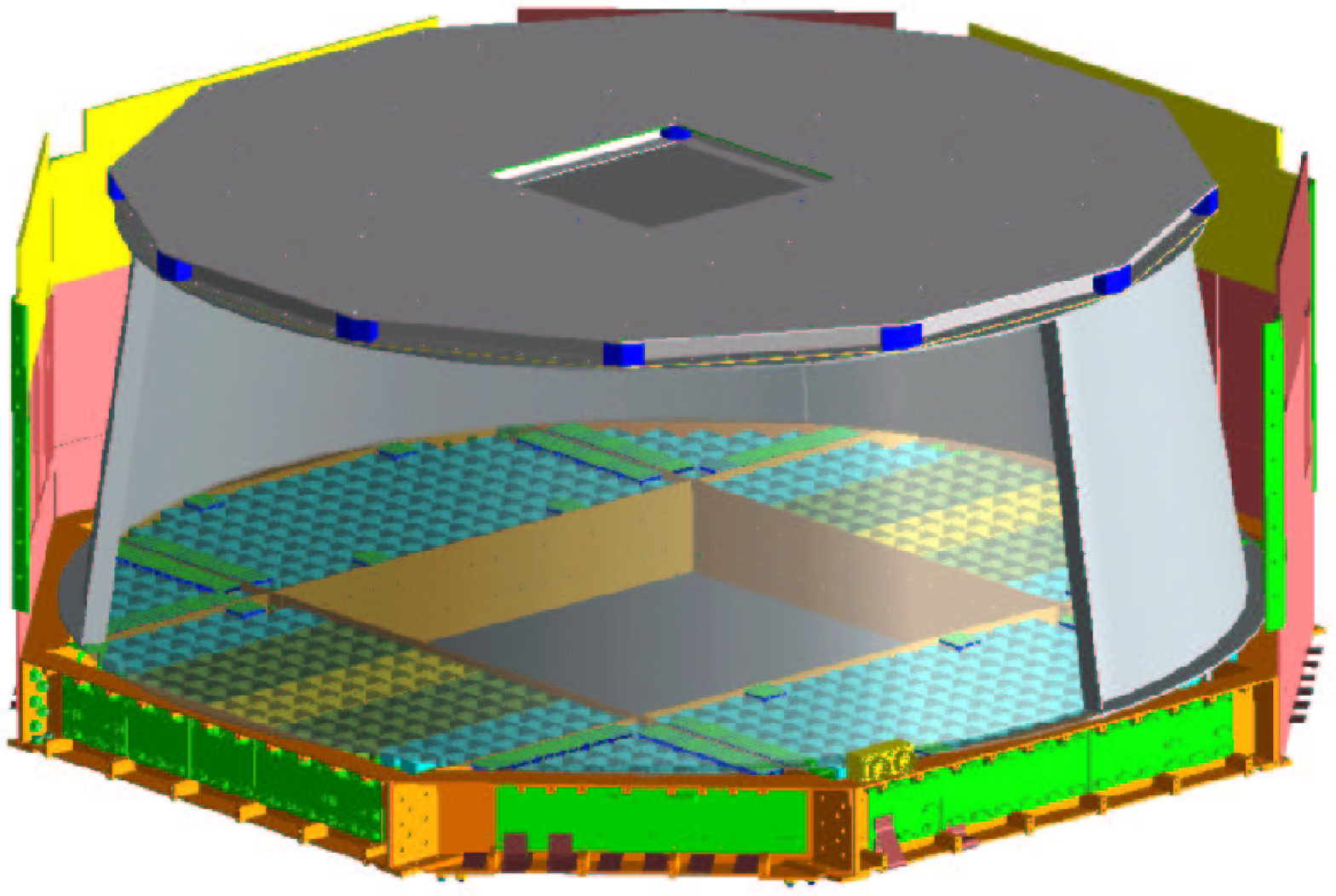}
&
\hspace{-.3cm}
\includegraphics[scale=0.3,angle=0,clip=,bb=14 14 496 463]{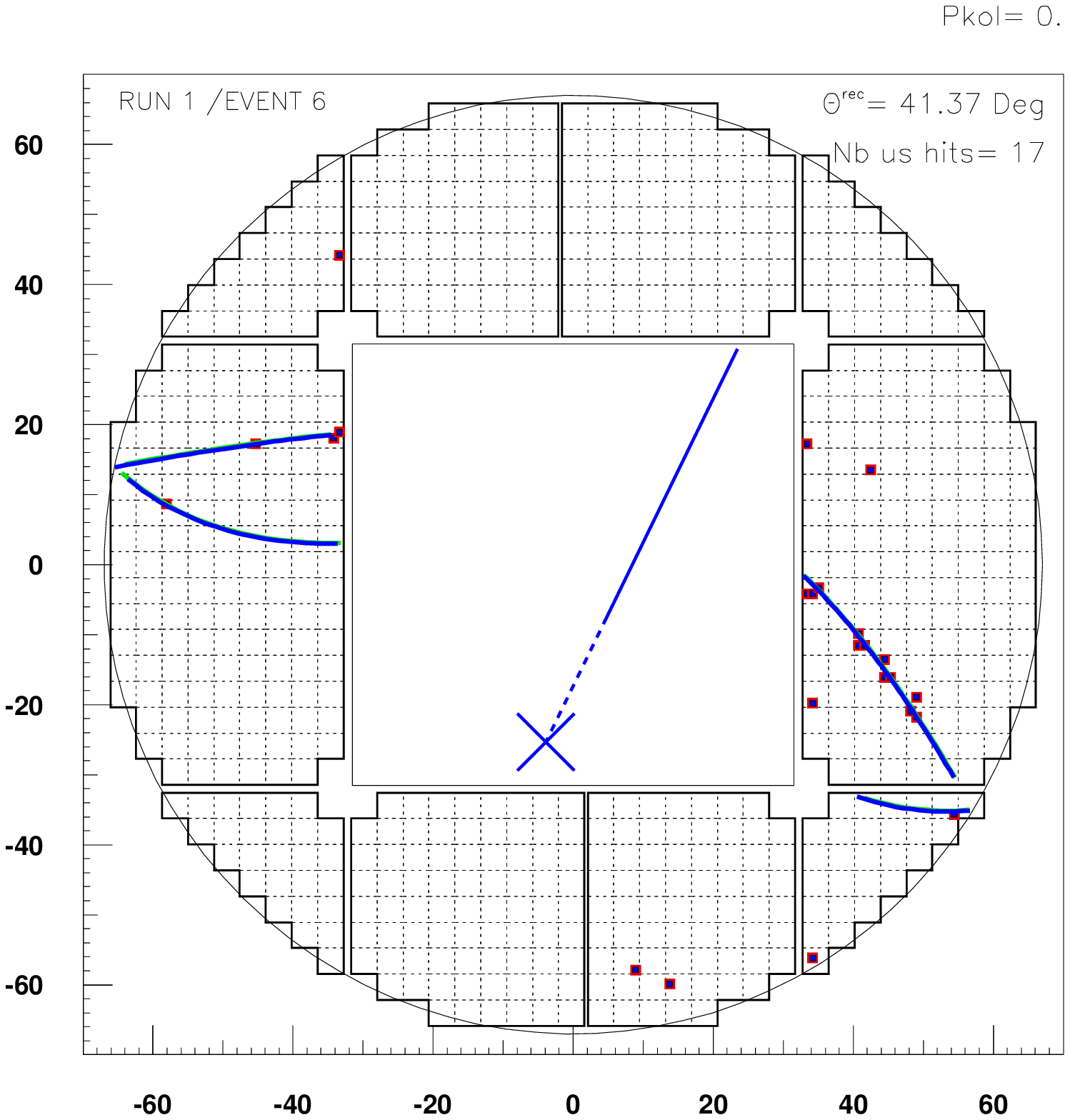} 
\end{tabular}
\caption{\emph{On the left:} View of the RICH detector. \emph{On the
  right:} Beryllium event display generated in a NaF radiator. The reconstructed photon pattern (full line)
  includes both reflected and non-reflected branches. The outer circular
  line corresponds to the lower 
boundary of the conical mirror. The square is the limit of the non-active
region.\label{fig:rich}}
\end{center} 
\end{figure}

The RICH of AMS is a proximity focusing \CK\ radiation detector.
Its radiator is composed by aerogel (n=1.05) and a sodium fluoride (NaF
1.334) squared region placed at the center and covering an acceptance of $\sim$10$\%$. The whole detector set
will be covered by a high reflectivity conical mirror increasing the 
reconstruction efficiency.  Photons will be detected in a 
matrix with 680 photomultipliers (PMTs) coupled to light guides. 
There will be a large non-active area at the center of the detection 
area due to the insertion of an electromagnetic calorimeter. 
For a more detailed description of the RICH detector see 
reference~[\refcite{bib:buenerd}].
Figure \ref{fig:rich} shows a view of the RICH and a beryllium event display
with a view of the PMT detailed matrix.

%\vspace{-0.8cm} 
\section{Velocity reconstruction}
A charged particle crossing a dielectric material of refractive index $n$,
with a velocity $\beta$, greater than the speed of light in that 
medium emits photons.
The aperture angle of the emitted photons with respect to the 
radiating particle track is known as the \CK\ angle, $\theta_c$,
and it is given by (see [\refcite{bib:rich}]):
%\vspace{-0.2cm}
\begin{equation}
\cos\theta_c = \frac{1}{\beta~n}
\label{eq1}
\end{equation}

It follows that the velocity of the particle, $\beta$, is straightforward
derived from the \CK\ angle reconstruction, which is based on a fit to the
pattern of the detected photons.
Complex photon patterns can occur at the detector plane 
due to mirror reflected photons, as can be seen on right display of Figure
\ref{fig:rich}. The event displayed is generated by a simulated beryllium nuclei in a
NaF radiator. 

The \CK\ angle reconstruction procedure 
relies on the information of the particle direction 
provided by the tracker.
The tagging of the hits signaling the passage 
of the particle through the solid light guides in the detection plane, 
provides an additional track ele\-ment, however, those hits are excluded from the reconstruction.
The best value of $\theta_c$ will result from the maximization of a likelihood function, 
built as the product of the probabilities, $p_i$, that the detected hits belong to a given (hypothesis)
\CK\ photon pattern ring,
%\vspace{-0.3cm}
\begin{equation}
L(\theta_c) = \prod_{i=1}^{nhits} p_i^{n_{i}}  \left[ r_i(\theta_c) \right].
\label{eq:likelihood}
%\vspace{-0.3cm}
\end{equation}

\begin{figure}[htb]
\begin{center}
\hspace{-.4cm}
\begin{tabular}{cc}
\hspace{-.5cm} 
\scalebox{0.38}{%                                                          
%\includegraphics[bb=20 -70 567 567]{arrudal/betaresol.vs.zscint.538.eps} 
%\begin{overpic}{arrudal/betaresol.vs.zscint.538.eps} 
\begin{overpic}[bb=0 10 482 482]{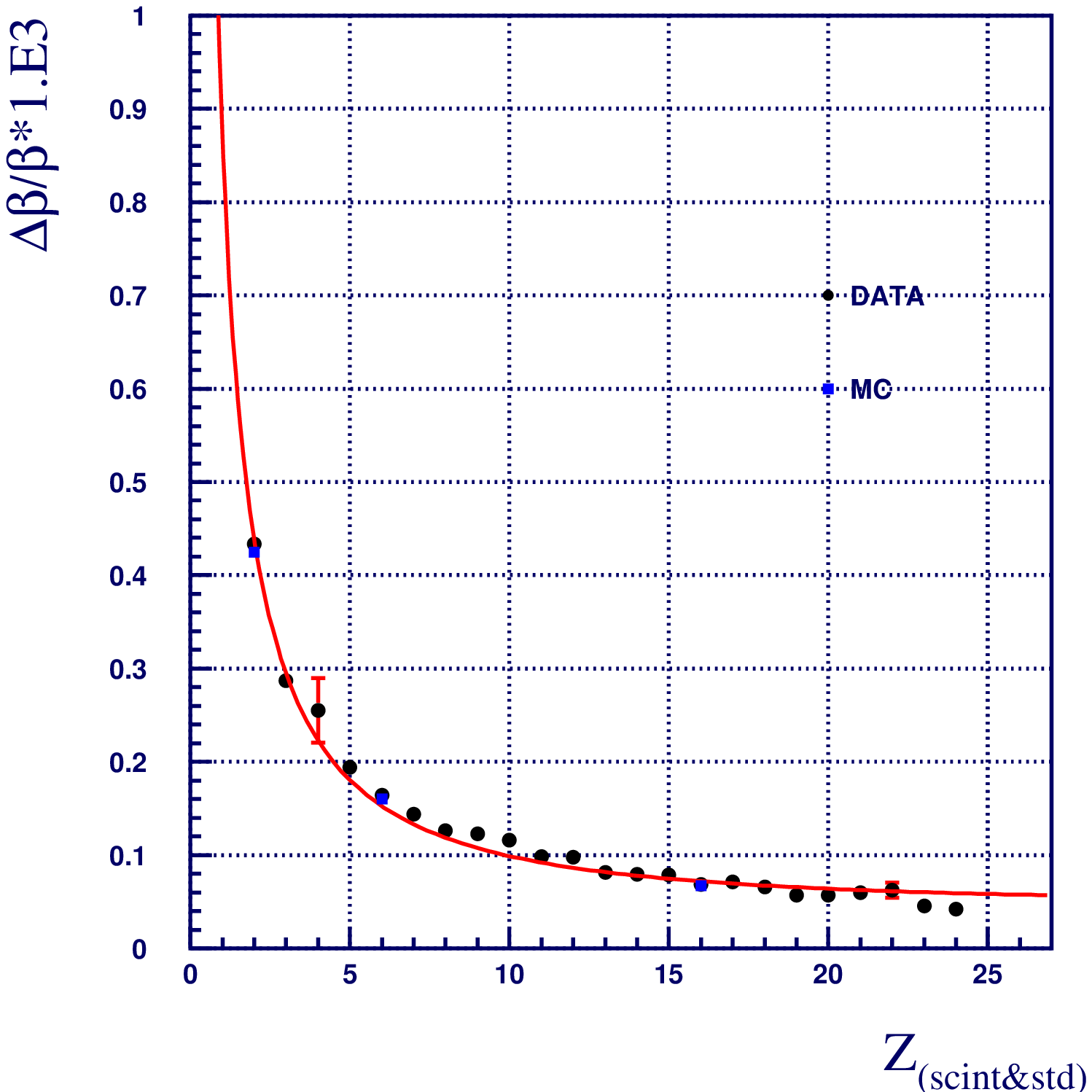}
\put(30,70){\Huge{$10^{-3}\sqrt {({\frac{A}{Z}})^{2}+B^{2}}$}}
\put(30,63){\Huge{$A=0.87\pm 0.003$}}
\put(30,56){\Huge{$B=0.047\pm 0.001$}}
\put(30,48){\Huge{$\chi^{2}=8.99/22$}}
\end{overpic}
}
&
\hspace{-0.7cm}
\scalebox{0.31}{%                                                          
\includegraphics[bb=0 -5 485 507]{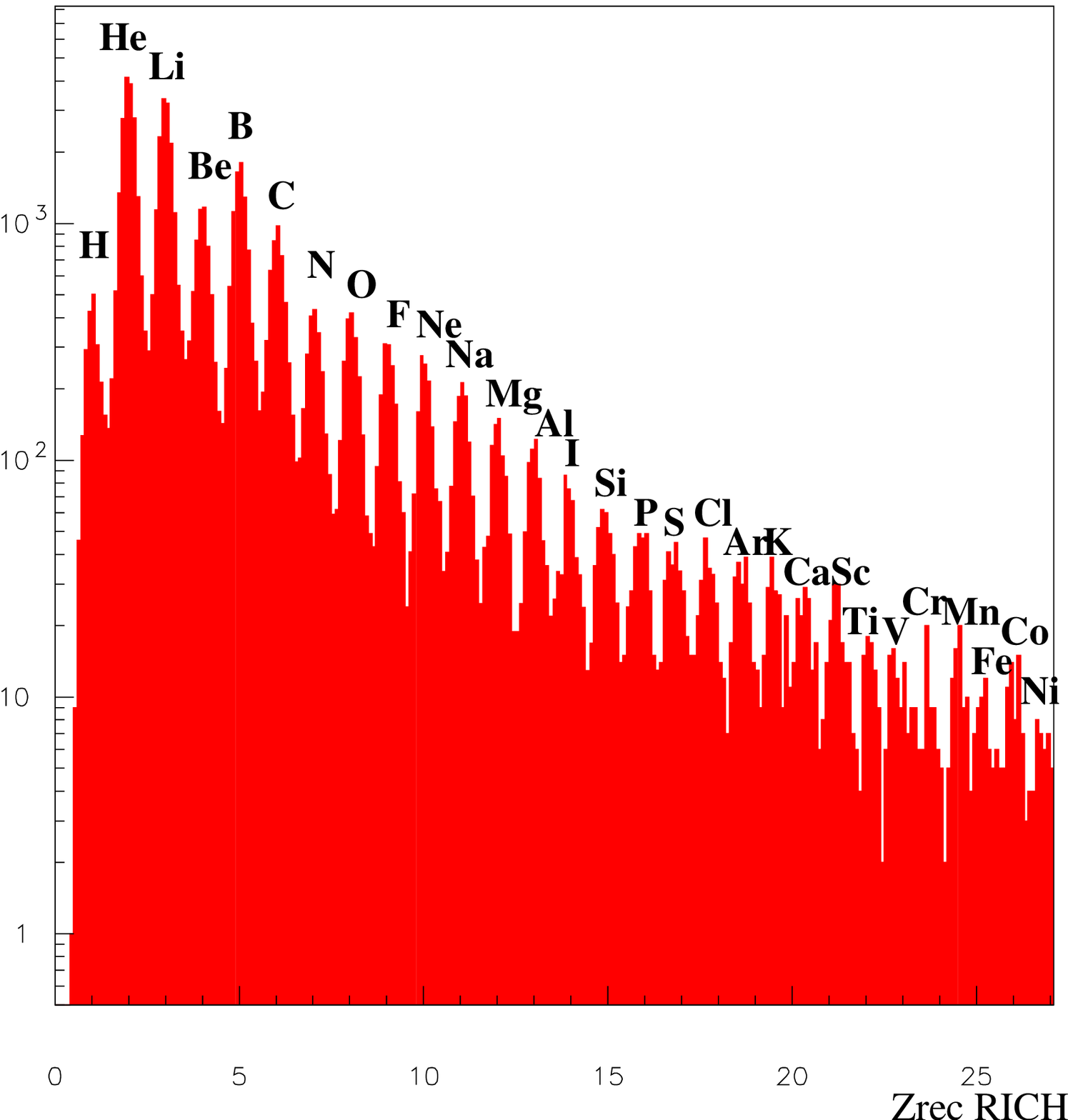} 
}                                                                          
\end{tabular} 
%\vspace{-0.3cm}
\caption{At left evolution of the relative resolution on beta with the
  charge and at right the reconstructed charge peaks. Both are
  reconstructions with data from a test beam at CERN in October 2003, using
  an indium beam of
158\,GeV/c/nuc. \label{fig:thc}}
\end{center}
\end{figure}

Here $r_i$ is the closest distance of the hit to the \CK\ pattern. For
a more complete description of the method see [\refcite{bib:NIM}].
The resolution achieved for protons of 20\,GeV/c/nuc is $\sim$4\,mrad. 
The evolution of the relative
resolution of beta with the charge can be observed on the left plot of
Figure \ref{fig:thc}. It was extracted from reconstructed events generated in
a test beam at CERN in October 2003 with fragments of an indium beam with a
momentum per nucleon of
158\,GeV/c/nuc, in a prototype of the RICH detector.
%\vspace{-0.8cm}

%\vspace{-0.8cm}
\section{Charge reconstruction}
The \CK\ photons produced in the radiator are uniformly emitted along the particle path 
inside the dielectric medium, $L$, and their number per unit of energy 
depends on the particle's charge, $Z$, and velocity, $\beta$, and on 
the refractive index, $n$, according to the expression:
\begin{equation}
\frac{dN_{\gamma}}{dE} \propto Z^2 L \left( 1-\frac{1}{\beta^2 n^2} \right)=
Z^2 L \sin^2\theta_c
\label{eq:dnde}
\end{equation}

So to reconstruct the charge the following procedure is required:

\begin{itemize}
%\vspace{-0.2cm}
\item \CK\ angle reconstruction. 
%\vspace{-0.3cm}
\item Estimation of the particle path, $L$, which relies on the information of the particle 
direction provided by the tracker.
%\end{itemize}
%\begin{itemize}
%\vspace{-0.3cm}
\item Counting the number of photoelectrons.\\
The number of photoelectrons related to the \CK\ ring has to be counted within a 
fiducial area, in order to exclude the uncorrelated background noise. Therefore, 
photons which are scattered in the radiator are excluded.  
A distance of 15 mm to the ring was defined as the limit for photoelectron
counting, corresponding to a ring width of $\sim$4 pixels.
%\vspace{-0.3cm}
\item Evaluation of the photon detection efficiency.
The number of radiated photons ($N_{\gamma}$) which will be detected ($n_{p.e.}$) 
is reduced due to 
the interactions with the radiator ($\varepsilon_{rad}$), the photon ring acceptance 
($\varepsilon_{geo}$), light guide ($\varepsilon_{lg}$) and photomultiplier efficiency 
($\varepsilon_{pmt}$).
\begin{equation}
n_{p.e.} \sim N_{\gamma}~\varepsilon_{rad}~\varepsilon_{geo}~\varepsilon_{lg}~\varepsilon_{pmt}  
\end{equation}
\end{itemize}

The charge is then calculated according to expression \ref{eq:dnde}, 
where the normalization constant can be evaluated from a calibrated beam of
charged particles.
In the right plot of Figure \ref{fig:thc} are visible reconstructed charge
peaks from the mentioned test beam at CERN in
October 2003. These results were obtained with aerogel radiator 1.05 and 2.5\,cm thick.
A charge resolution for helium events  slightly  better than $\Delta Z \sim 0.2$ was observed together with a systematic of $1\%$. A clear charge separation up to Z=27 was achieved.
For a more complete description of the charge reconstruction method see [\refcite{bib:NIM}].
\section{Isotopic Element Separation}
Isotopic separation and particularly the ratios $^3$He/$^4$He and
 $^{10}$Be/$^9$Be is a major part of the physics goals where the RICH plays a
 fundamental role within AMS. The presence of a mixed
radiator with a NaF radiator at the center
will allow AMS to cover a kinematic energy range from 0.5\,GeV/nucleon up to
 around 10\,GeV/nucleon.

Samples of helium and beryllium nuclei corresponding to 1 day and 1 year of
 data taking, respectively, were simulated. These samples were generated
 according to [\refcite{bib:Seo}] for helium and [\refcite{bib:Stg&Mosk}] for
 beryllium nuclei. Afterwards, the spectra was modulated taking into account
 the geomagnetic field. The masses were reconstructed using a momentum uncertainty $\sim$2$\%$.

The reconstructed masses were fitted with a sum of two gaussian functions:
$$
%f(m) = \frac{N_1}{\sigma_1\sqrt{2\pi}} \exp{-\frac{1}{2}\left(\frac{m-M_1}{\sigma_1}\right)^2} +
%       \frac{N_2}{\sigma_2\sqrt{2\pi}}
%       \exp{-\frac{1}{2}\left(\frac{m-M_2}{\sigma_2}\right)^2} 
f(m) \propto \alpha\left( G_1(M_1,\sigma_1)+G_2(M_2,\sigma_2) \right)
%       \frac{N_2}{\sigma_2\sqrt{2\pi}} \exp{-\frac{1}{2}\left(\frac{m-M_2}{\sigma_2}\right)^2} 
$$  
where $M_i$, $\sigma_i$ and $\alpha$ are respectively the isotopic mass
central value, the mass width and the relative weight of the two distributions.

\begin{figure}
\begin{center}
\vspace{-0.8cm}
\begin{tabular}{cc}   
%\vspace{-0.1cm}
\hspace{-0.5cm}
\scalebox{.34}{\includegraphics[bb=0 0 454 454]{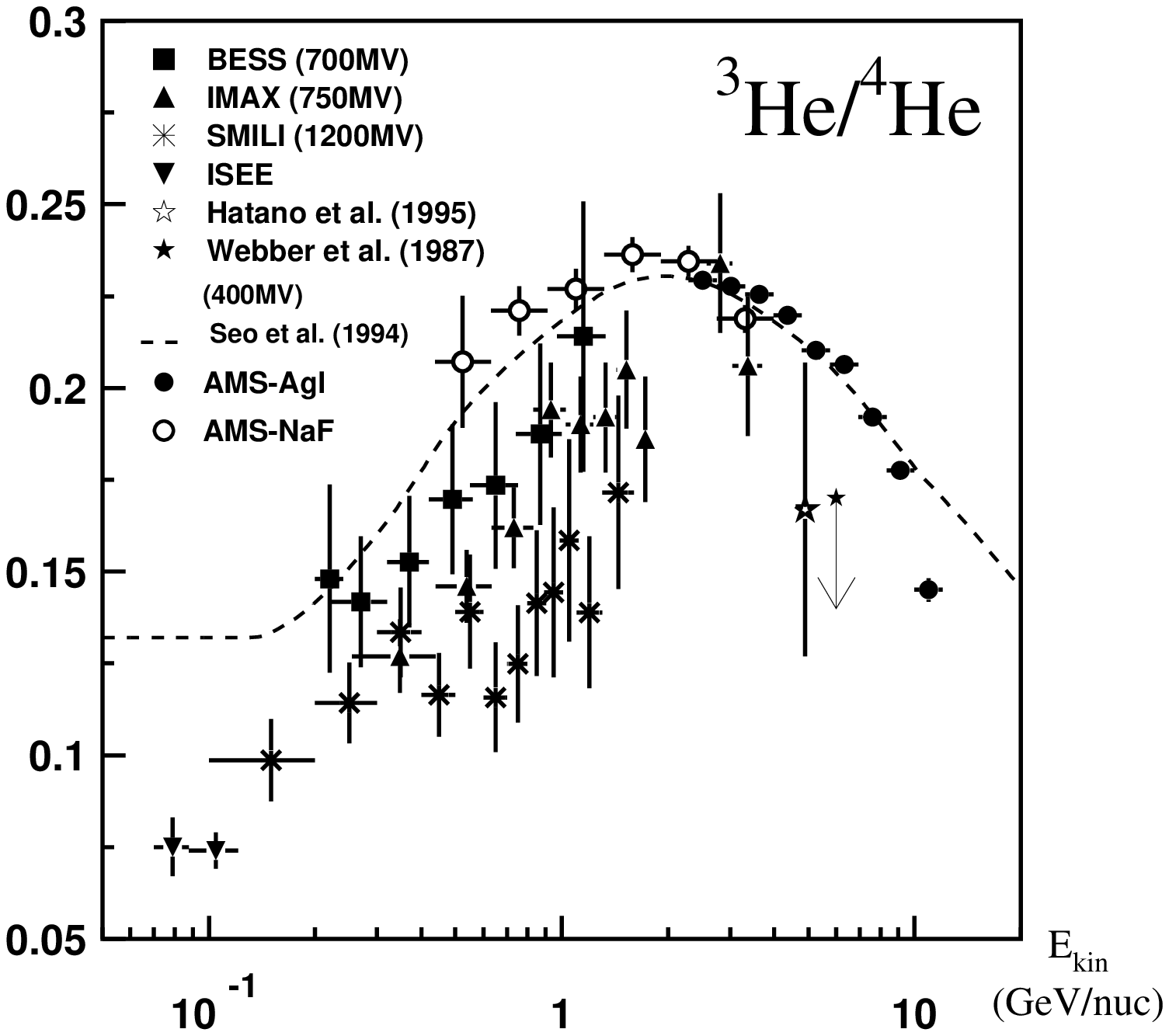}}
&
%\hspace{-1.6cm}
\scalebox{.34}{\includegraphics*{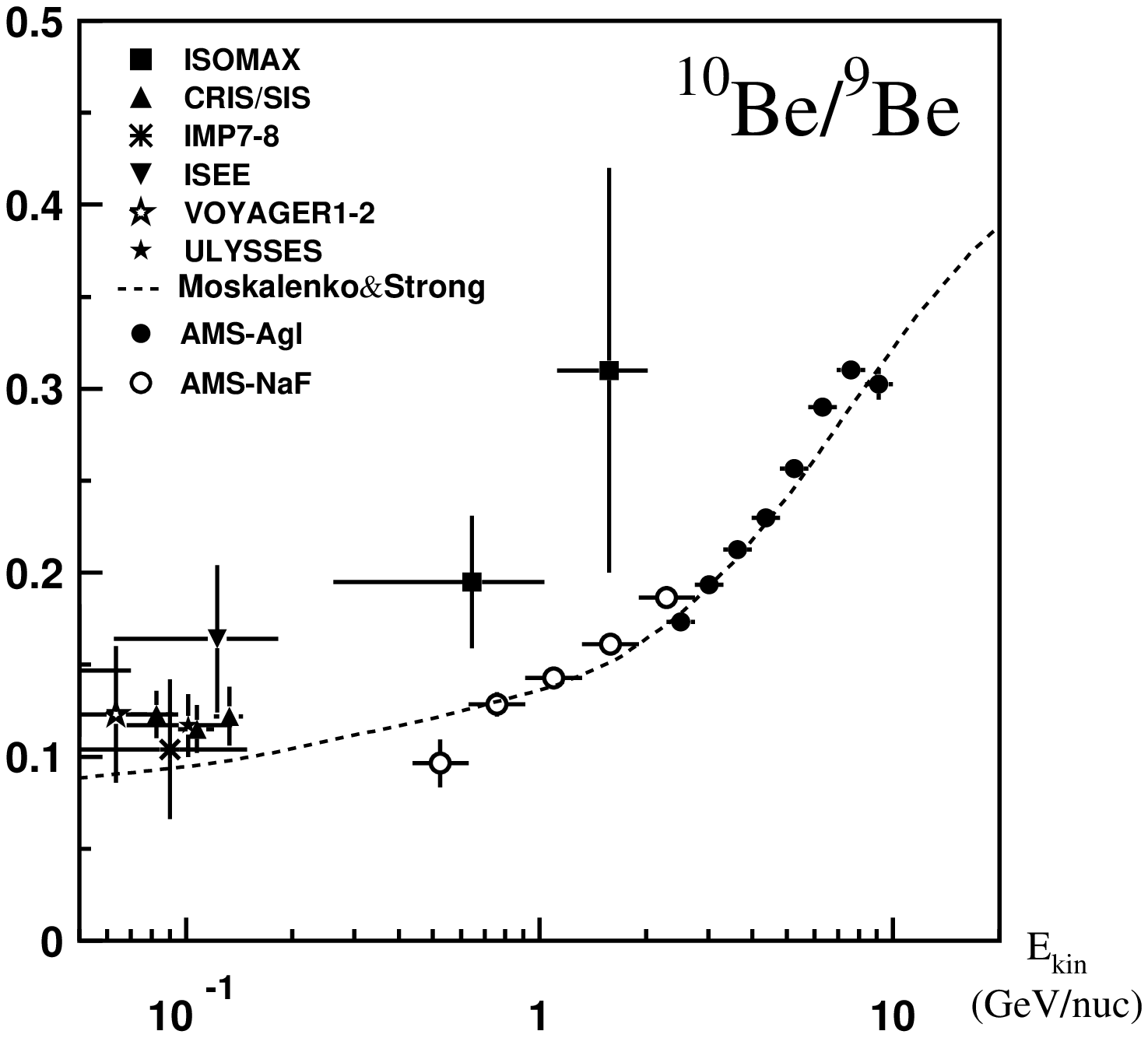}}
\end{tabular}
%\vspace{-0.3cm}
\caption{Reconstructed isotopic ratios of helium and beryllium simulated events as function of kinetic energy per nucleon. The
aerogel in study has a refractive index of 1.050. \label{fig:mass}}
\end{center}
\end{figure}

Figure \ref{fig:mass} presents the isotopic ratios obtained from the fits as function of the 
kinetic energy.
Isotopic ratios from events crossing the sodium fluoride radiator 
are fairly measured up to the aerogel threshold.
From there on, the aerogel allows to measure the isotopic ratios up to
around 10\,GeV/nucleon of kinetic energy. Above 10\,GeV/nuc the mass
relative resolution is greater than 8.5\% for He and greater than 6\% for Be.
%
%\vspace{-.8cm}
\section{Conclusions}
AMS is a spectrometer designed for antimatter, dark matter searches and for measuring
relative abundances of nuclei and isotopes.
The instrument will be equipped with a proximity focusing RICH detector based
on a mixed radiator of aerogel and sodium fluoride, enabling velocity measurements with a resolution of about 0.1\% and extending the charge measurements up to the iron element.
Velocity reconstruction is made with a likelihood method. Charge reconstruction is made in an event-by-event basis. 
Both algorithms were successfully applied to simulated data samples with flight configuration.
Evaluation of the algorithms on real data taken with the RICH prototype was
performed at the LPSC, Grenoble in 2001 and in the test beam at CERN, in October 2002 and 2003.
The RICH radiator will allow AMS to perform helium and beryllium isotopic separation up to 10\,GeV/nucleon.

%\bibliography{biblio}
%\begin{chapthebibliography}{1}                                 

%\end{chapthebibliography}

\end{document}